\documentclass[preprint, pre,superscriptaddress, letterpaper, fleqn, floatfix, showpacs ]{revtex4}

\usepackage{euscript, units, amsfonts,amsmath, graphics, graphicx, dcolumn, fancyhdr}
\usepackage{times}

\usepackage[usenames]{color}

\usepackage{amssymb}

\begin{document}

\title{Brownian Dynamics of charged particles in a constant magnetic field}

\author{L. J. Hou}\email{ljhouwang@gmail.com}
\affiliation{IEAP, Christian-Albrechts Universit\"{a}t, D-24118 Kiel, Germany}

\author{Z.\ L.\ Mi\v{s}kovi\'{c}}
\affiliation{Department of Applied Mathematics, University of Waterloo, Waterloo, Ontario, Canada N2L 3G1}
\author{A. Piel}
\affiliation{IEAP, Christian-Albrechts Universit\"{a}t, D-24118 Kiel, Germany}
\author{P. K. Shukla}
\affiliation{Institut f\"{u}r Theoretische Physik IV, Ruhr-Universit\"{a}t Bochum, D-44780, Germany}

\begin{abstract}
Numerical algorithms are proposed for simulating the Brownian dynamics of charged particles in an external magnetic field,
taking into account the Brownian motion of charged particles, damping effect and the effect of magnetic field self-consistently.
Performance of these algorithms is tested in terms of their accuracy and long-time stability by using a three-dimensional
Brownian oscillator model with constant magnetic field. Step-by-step recipes for implementing these algorithms are given in
detail. It is expected that these algorithms can be directly used to study particle dynamics in various dispersed systems in the
presence of a magnetic field, including polymer solutions, colloidal suspensions and, particularly complex (dusty) plasmas. The
proposed algorithms can also be used as thermostat in the usual molecular dynamics simulation in the presence of magnetic field.
\end{abstract}

 \maketitle

\section{Introduction}\label{Sec_intro}
Brownian Dynamics (BD) simulation method for many-body systems of particles immersed in a liquid, gaseous or plasma medium
\cite{Ermak,Allen,vanGunsteren1982,Branka,Chin,Allen1989,Lemons1999,Lemons2002} can be regarded as a generalization of the usual
Molecular Dynamics (MD) method for many-body systems in free space. While the MD method is based on Newton's equations of
motion, the BD method is based on their generalization in the form of Langevin equation and its integral \cite{Lemons2002}:
\begin{eqnarray}
\frac{d}{dt}\mathbf{v}&=&-\gamma
\mathbf{v}+\frac{1}{m}\mathbf{F}+\mathbf{A}(t) ,\label{eqlangevin0} \\
\frac{d}{dt}\mathbf{r}&=&\mathbf{v} \nonumber
\end{eqnarray}
where, as usual, $m$, $\mathbf{v}$ and $\mathbf{r}$ are, respectively, the mass, velocity and position of a Brownian particle,
whereas $\mathbf{F}$ is a systematic (deterministic) force coming from external fields and/or from inter-particle interactions
within the system. What is different from Newton's equations is the appearances of dynamical friction, $-\gamma \mathbf{v}$, and
random, or Brownian acceleration, $\mathbf{A}(t)$. These two force components represent two complementing effects of a single,
sub-scale phenomenon: numerous, frequent collisions of the Brownian particle with molecules in the surrounding medium. While the
friction represents an average effect of these collisions, the random acceleration represents fluctuations due to discreteness
of collisions with molecules, and is generally assumed to be a delta-correlated Gaussian white noise. The friction and random
acceleration are related through a fluctuation-dissipation theorem which includes the ambient temperature, therefore
guaranteeing that a Brownian particle can ultimately reach thermal equilibrium within the medium \cite{Lemons2002}.

The Langevin equations, Eq.\ (\ref{eqlangevin0}), can be numerically integrated in a manner similar to Newton's equations in the
MD simulation, which gives rise to several algorithms for performing BD simulation, such as the Euler-like \cite{Ermak},
Beeman-like \cite{Allen,Allen1989}, Verlet-like \cite {vanGunsteren1982}, and Gear-like Predictor-Corrector (PC) methods
\cite{Hou2008pop}, as well as a wide class of Runge-Kutta-like algorithms (see, e.g., \cite{Branka,Chin}). All those methods
were used successfully to study problems in various dispersed systems, such as polymer solutions \cite{Ottinger}, colloidal
suspensions \cite{Chen2004} and, in particular, complex (dusty) plasmas
\cite{Hou2008pop,Zheng1995,Hoffmann2000,Vaulina2002,Hou2006,Hou2009}.

Recently, there has been a growing interest in studying the dynamics of dust particles and dust clouds in both unmagnetized
\cite{Donko2009,Mendonca1997} and magnetized dusty \cite{Amatucci2004,Sato2001,Kaw2002,Shukla2002} plasmas. The topics studied
so far include, besides Brownian motion of a dust particle in an unmagnetized plasma \cite{Mendonca1997}, also the gyromotion of
a single dust particle and rotations \cite{Amatucci2004} of dust clouds \cite{Sato2001,Kaw2002,Shukla2002} and clusters
\cite{Juan1998,Konopka2000,Cheung2003,Carsternsen2009} in a magnetized plasma. There have also been several theoretical
proposals for studying waves and collective dynamics \cite{Shukla2009} in such magnetized plasma systems, in which dust
particles are fully magnetized \cite{Uchida2004,Jiang2007,Jiang2007pop,Feldmann2008,Farokhi2009}. However, exploring those
proposals in the laboratory does not seem to be quite feasible as yet, due to many constrains
\cite{Uchida2004,Jiang2007,Jiang2007pop,Farokhi2009}. Therefore, it is desirable to have algorithms for numerical experiments
that can validate the existing theories, on one hand, and that can serve as a guide for future laboratory experiments, on the
other hand. Since there are no such algorithms, to the best of our knowledge, we propose here a few new BD algorithms, which
treat an external magnetic field in the simulation in a manner consistent with the Langevin dynamics, Eq.\ (\ref{eqlangevin0}).

The manuscript is organized in the following fashion. In Sec.\ \ref{Sec_gen}, we present the general formula for a BD simulation
with a constant magnetic field. Detailed implementations to the Euler-, Beeman- and Gear-like methods are given, respectively,
in Sections \ref{Sec_EL}, \ref{Sec_BL}, and \ref{Sec_GL}. Concluding remarks are contained in Sec.\ \ref{Sec_con}.

\section{General formula for BD simulations}\label{Sec_gen}

The dynamics of a charged Brownian particle in a constant external magnetic field $\mathbf{B}$ is described by introducing the
Lorentz force in the Langevin equation \cite{Lemons1999,Lemons2002}

\begin{eqnarray}
\frac{d}{dt}\mathbf{v}&=&-\gamma
\mathbf{v}+\frac{1}{m}\mathbf{F}+\frac{Q}{mc}\,\mathbf{\mathbf{v}\times\mathbf{B}}+\mathbf{A}(t) ,
\nonumber \\
\frac{d}{dt}\mathbf{r}&=&\mathbf{v}  \label{eqlangevin},
\end{eqnarray}
where $Q$ is the charge on the particle and $c$ is the speed of light in vacuum.

As usual, certain assumptions must be made about the deterministic force $\mathbf{F}$ in order to construct a meaningful
algorithm for a many-particle simulation from the above equation. A common approach
\cite{Ermak,Allen,vanGunsteren1982,Allen1989} is to assume that $\mathbf{F}$ is only an explicit function of time $t$. Thus, a
Taylor series of $\mathbf{F}$ or, equivalently the deterministic acceleration, $\mathbf{a}\equiv\mathbf{F}/m$, can be written as

\begin{equation}
\mathbf{a}(t) = \mathbf{a}(0) + \dot{\mathbf{a}}(0)t + \frac{1}{2!}\ddot{\mathbf{a}}(0)t^2 +
\frac{1}{3!}\dddot{\mathbf{a}}(0)t^3 + \cdots + \frac{1}{n!}\mathbf{a}^{(n)}(0)t^n + \cdots, \label{eqtaylor}
\end{equation}
where $\mathbf{a}^{(n)}$ represents the $n$th-order time derivative of $\mathbf{a}$. There are various ways to derive formulas
for conducting a BD simulation from the Langevin equation, Eq.\ (\ref{eqlangevin}), based on the above Taylor series. We shall
adopt the strategy outlined in Refs.\ \cite{Ermak,Allen,vanGunsteren1982,Allen1989}, and more recently implemented in Refs.\
\cite{Lemons1999,Lemons2002}, as it is simple and straightforward, especially for readers with some simulation background but
without much background, or interest in stochastic calculus.

The Langevin equations, Eq.\ (\ref{eqlangevin}), may be integrated analytically in a short time, based on an adopted truncation
rule for the series in Eq.\ (\ref{eqtaylor}), thereby giving an updating formula for a BD simulation. (We note that detailed
technique for integrating the Langevin equation is particularly well described in Refs.\ \cite{Lemons1999,Lemons2002}.) The
resultant formulas are actually expressions for the two random variables, $\mathbf{v}(t)$ and $\mathbf{r}(t)$, which, under the
assumptions that the Brownian acceleration in Eq.\ (\ref{eqlangevin}) is a Gaussian white noise and that $\gamma$ is constant,
turn out to be normally distributed random variables, according to the \emph{normal linear transform theorem}
\cite{Lemons1999,Lemons2002}. Therefore, the Cartesian components of the velocity $\mathbf{v}=\{v_x,v_y,v_z\}$ and position
$\mathbf{r}=\{x,y,z\}$ vectors for a Brownian particle can be expressed in terms of their respective means and variances

\begin{eqnarray}
v_\alpha(t) &=& \text{mean}\{v_\alpha(t)\}+ \sqrt{\text{var}\{v_\alpha(t)\}}\,N_\alpha^\mathbf{v}(0,1), \nonumber \\
\alpha(t) &=& \text{mean}\{\alpha(t)\}+ \sqrt{\text{var}\{\alpha(t)\}}\,N_\alpha^\mathbf{r}(0,1), \label{eqnormal}
\end{eqnarray}
where $\alpha$ takes values $x$, $y$ and $z$, and $N(0,1)$ is a shorthand notation for the standard normal random variable
having zero mean and unit variance, the so-called \emph{unit normal} \cite{Lemons1999,Lemons2002}. The superscripts attached to
the components of random vectors $\mathbf{N}^\mathbf{v}=\{N_x^\mathbf{v},N_y^\mathbf{v},N_z^\mathbf{v}\}$ and
$\mathbf{N}^\mathbf{r}=\{N_x^\mathbf{r},N_y^\mathbf{r},N_z^\mathbf{r}\}$, appearing in Eq.\ (\ref{eqnormal}), indicate that
those two sets of unit normals are associated, respectively, with the velocity and position of the Brownian particle. We note
that the Cartesian components of the vector $\mathbf{N}^\mathbf{v}$ are mutually independent, as are the components of the
vector $\mathbf{N}^\mathbf{r}$, but the vectors $\mathbf{N}^\mathbf{v}$ and $\mathbf{N}^\mathbf{r}$ are correlated, as will be
shown below.

We should emphasize the importance of Eq.\ (\ref{eqnormal}) because it provides a general updating formula for the BD simulation
and also enables substantial simplifications in the subsequent design of our algorithm. Now, solving the stochastic differential
equation, Eq.\ (\ref{eqlangevin}), and obtaining the two random variables $\mathbf{v}(t)$ and $\mathbf{r}(t)$, is simply reduced
to determining the two sets of deterministic quantities, the means and variances of $\mathbf{v}(t)$ and $\mathbf{r}(t)$, as well
as their covariances, assuming an appropriate truncation scheme for the deterministic acceleration, Eq.\ (\ref{eqtaylor}). To
simplify the notation, in the following we shall denote the velocity and position means, respectively, by $\langle
v_\alpha\rangle\equiv\text{mean}\{v_\alpha\}$ and $\langle\alpha\rangle\equiv\text{mean}\{\alpha\}$, and we shall use standard
deviations $\sigma$ instead of variances.

\subsection{Variances and covariances}

We begin with variances and covariances because they do not depend on the form of deterministic acceleration, $\mathbf{a}(t)$.

It should be noted that random increments in the velocity do not depend explicitly on the magnetic field and are therefore
isotropic. The corresponding variances are then given by $\text{var}\{ v_x\}=\text{var}\{ v_y\}=\text{var}\{ v_z\}=\text{var}\{
v\}=\sigma_{v}^2$, where

\begin{equation}
\sigma_{v}=\sqrt{\frac{k_BT}{m}(1-e^{-2\gamma t})}, \label{eqvarv}
\end{equation}
with $k_B$ being the Boltzmann constant and $T$ the temperature of the medium. [It should be noted that in an equilibrium
between the Brownian particle and the medium, $T$ is also the kinetic temperature of the Brownian particle. However, Brownian
particles may have kinetic temperature that is different from $T$, which opens a possibility of simulating non-equilibrium
processes by using the BD].

However, random displacements of the position do depend explicitly on the magnetic field and therefore are non-isotropic. Let us
assume $\mathbf{B}=\{0,0,B\}$ and define $\Omega=QB/(cm)$ to be the gyrofrequency of a Brownian charged dust particle. Then, we
have $\text{var}\{ x\}=\text{var}\{y\}=\sigma^2_{\perp}$ and $\text{var}\{z\}=\sigma^2_{\parallel}$, where $\sigma_{\perp}$ and
$\sigma_{\parallel}$ are, respectively, standard deviations of the position in the directions perpendicular and parallel to the
external magnetic field \cite{Lemons1999,Lemons2002}. In particular, we find

\begin{eqnarray}
\sigma_{\perp}&=&t\sqrt{\frac{\gamma^2}{\gamma^2+\Omega^2} \frac{2kT}{m} \frac{1}{\gamma t} \left[ 1+\frac{1-e^{-2\gamma
t}}{2\gamma t}-\frac{2\gamma^2}{\gamma^2+\Omega^2}\frac{1-e^{-\gamma t}
(\cos\Omega t-\frac{\Omega}{\gamma} \sin\Omega t)}{\gamma t} \right]}, \nonumber \\
\sigma_{\parallel}&=&t\sqrt{\frac{2kT}{m\gamma t}\left(1-2\frac{1-e^{-\gamma t}}{\gamma t}+\frac{1-e^{-2\gamma t }} {2\gamma
t}\right)}. \label{eqvarr}
\end{eqnarray}

Furthermore, we have $(6\times 5)/2$ covariances \cite{Lemons1999,Lemons2002}, most of which are zeros. The non-zero covariances
are $\text{cov}\{v_x,x\}$, $\text{cov}\{v_y,y\}$, $\text{cov}\{v_x,y\}$, $\text{cov}\{v_y,x\}$ and $\text{cov}\{v_z,z\}$
\cite{Lemons1999,Lemons2002} (here, we use the definition $\text{cov}\{X,Y\}=\langle XY\rangle-\langle X\rangle \langle
Y\rangle$ for the covariance of two random variables, $X$ and $Y$). In particular, we obtain \cite{Lemons1999,Lemons2002}

\begin{eqnarray}
&K&\equiv\text{cov}\{ x,v_x\}=\text{cov}\{ y,v_y\}=t\frac{kT}{m}\frac{1}{\gamma t}
\frac{\gamma^2}{\gamma^2+\Omega^2}\left( 1+e^{-2\gamma t}-2e^{-\gamma t}\cos \Omega t\right), \nonumber \\
&H&\equiv\text{cov}\{ y,v_x\}=-\text{cov}\{ x,v_y\}=t\frac{kT}{m} \frac{1}{\gamma t}
\frac{\gamma\Omega}{\gamma^2+\Omega^2}\left( 1-e^{-2\gamma t}-2e^{-\gamma t}\frac{\gamma}{\Omega}
\sin \Omega t \right), \nonumber \\
&L&\equiv\text{cov}\{v_z,z\}={t\frac{kT}{m\gamma t}\left(1-2e^{-\gamma t}+e^{-2\gamma t} \right )}. \label{eqcov}
\end{eqnarray}
We note that, when $B\rightarrow 0$, i.e., $\Omega\rightarrow 0$, one has $H\rightarrow 0$ and $K\rightarrow L$, so that
$\sigma_{\perp}\rightarrow \sigma_{\parallel}$, recovering the results for systems without magnetic field \cite{Lemons2002}.

\subsection{Mean values}

Given the expression for $\mathbf{a}(t)$ in terms its Taylor series in Eq.\ (\ref{eqtaylor}), one can obtain the means $\langle
\mathbf{v}\rangle=\{\langle v_x\rangle,\langle v_y\rangle,\langle v_z\rangle\}$ and $\langle \mathbf{r}\rangle=\{\langle
x\rangle,\langle y\rangle,\langle z\rangle\}$ in a number of ways by integrating Eq.\ (\ref{eqlangevin}). For simplicity, we
follow here Lemons' methodology \cite{Lemons1999,Lemons2002}, in which Eq.\ (\ref{eqlangevin}) is reduced to a set of
deterministic (ordinary) differential equations by taking expectation values of both sides
\begin{eqnarray}
\frac{d\langle\mathbf{v}\rangle}{dt} &=& -\gamma\langle\mathbf{v}\rangle + \mathbf{a}(t)
+ \frac{Q}{mc}\langle\mathbf{\mathbf{v}\rangle\times\mathbf{B}},\label{eqmean} \\
\frac{d\langle\mathbf{r}\rangle}{dt} &=& \langle\mathbf{v}\rangle,  \nonumber
\end{eqnarray}
and using the fact that the Brownian acceleration $\mathbf{A}(t)$ is a Gaussian white noise with mean $\mathbf{0}$.

Equation \ (\ref{eqmean}) can be integrated analytically by assuming that initial conditions $\mathbf{r}_{0}$, $\mathbf{v}_{0}$,
$\mathbf{a}_{0}$, $\mathbf{\dot{a}}_{0}$, $\mathbf{\ddot{a}}_{0}$, $\mathbf{\dddot{a}}_{0}$, $\ldots, \mathbf{a}_{0}^{(n)}$, are
known at $t=0$. One ends up with
\begin{eqnarray}
\langle\mathbf{r}\rangle&=&\mathbf{r}_{0} + \mathbf{I}_{1}\cdot\mathbf{v}_{0}t + \mathbf{I}_{2}\cdot\mathbf{a}_{0}t^2 +
\mathbf{I}_{3}\cdot\mathbf{\dot{a}}_{0}t^3 + \cdots + \mathbf{I}_{n}\cdot\mathbf{a}_{0}^{(n-2)}t^n +
\cdots, \nonumber \\
\langle\mathbf{v}\rangle&=&\mathbf{I}_{0}\cdot\mathbf{v}_{0}+ \mathbf{I}_{1}\cdot\mathbf{a}_{0}t +
\mathbf{I}_{2}\cdot\mathbf{\dot{a}}_{0}t^2 + \mathbf{I}_{3}\cdot \mathbf{\ddot{a}}_{0}t^3 + \cdots +
\mathbf{I}_{n}\cdot\mathbf{a}_{0}^{(n-1)}t^{n} + \cdots. \label{eqglmean}
\end{eqnarray}
With the assumption $\mathbf{B}=\{0,0,B\}$, the matrices $\mathbf{I}_{n}$ are given by
\begin{equation}
\mathbf{I}_{n}= \left [
\begin{matrix}
  c_{n} &  b_{n}  & 0\\
  -b_{n} & c_{n} & 0\\
  0 &  0 & d_{n}\\
\end{matrix}
\right ],
\end{equation}
where
\begin{eqnarray}
b_{0} &=&e^{-\gamma t}\sin{\Omega t} \nonumber \\
c_{0} &=&e^{-\gamma t}\cos{\Omega t} \nonumber \\
d_{0} &=&e^{-\gamma t},
\end{eqnarray}
for $n=0$, and
\begin{eqnarray}
b_{n}&=&\frac{\Omega\gamma}{\gamma^2+\Omega^2}\frac{1}{(n-1)!\gamma t}
\left [1-(n-1)!c_{n-1}-\frac{\gamma}{\Omega}(n-1)!b_{n-1} \right ]  \nonumber \\
c_{n}&=&\frac{\gamma^2}{\gamma^2+\Omega^2}\frac{1}{(n-1)!\gamma t}
\left [1-(n-1)!c_{n-1}+\frac{\Omega}{\gamma}(n-1)!b_{n-1} \right ] \nonumber \\
d_{n}&=&\frac{1}{(n-1)!\gamma t}\left[1-(n-1)!d_{n-1}\right],  \label{eqcn}
\end{eqnarray}
for $n\geq 1$.

It is interesting to note that the coefficients satisfy very simple recursive relations
\begin{eqnarray}
\frac{d}{dt}(b_{n}t^{n}) = b_{n-1}t^{n-1}, \nonumber \\
\frac{d}{dt}(c_{n}t^{n}) = c_{n-1}t^{n-1}, \nonumber \\
\frac{d}{dt}(d_{n}t^{n}) = d_{n-1}t^{n-1}.
\end{eqnarray}
A special yet important case would be that of $B=0$, i.e. $\Omega=0$, and consequently $b_{n}=0$ and $c_{n}=d_{n}$, so that
$\mathbf{I}_{n}$ becomes a diagonal matrix with the diagonal elements being $d_{n}$ and, Eq. (\ref{eqglmean}) reads as
\cite{Hou2008pop}
\begin{eqnarray}
\langle\mathbf{r}\rangle_{B=0}&=&\mathbf{r}_{0} + d_{1}\mathbf{v}_{0}t + d_{2}\mathbf{a}_{0}t^2 + d_{3}\mathbf{\dot{a}}_{0}t^3 +
\cdots + d_{n}\mathbf{a}_{0}^{(n-2)}t^n +
\cdots, \nonumber \\
\langle\mathbf{v}\rangle_{B=0}&=&d_{0}\mathbf{v}_{0}+ d_{1}\mathbf{a}_{0}t + d_{2}\mathbf{\dot{a}}_{0}t^2 + d_{3}
\mathbf{\ddot{a}}_{0}t^3 + \cdots + d_{n} \mathbf{a}_{0}^{(n-1)}t^{n} + \cdots. \label{eqglmean_B0}
\end{eqnarray}
Further, when $\gamma\rightarrow 0$, $c_{n}$ and $d_{n}$ reduce to the usual coefficients of a Taylor series.

\subsection{General updating formula}
With the variances, covariances and means known, the updating formula for a BD simulation can be further written in a compact
form as \cite{Lemons1999,Lemons2002}
\begin{eqnarray}
\mathbf{v}(t) & = & \langle\mathbf{v}\rangle + \sigma_{v}\mathbf{N_{1}}(0,1), \nonumber \\
\mathbf{r}(t) & = & \langle\mathbf{r}\rangle + \mathbf{I}\cdot\mathbf{N_{1}}(0,1) + \mathbf{J}\cdot\mathbf{N_{2}}(0,1),
\label{equpdate}
\end{eqnarray}
where
\begin{equation}
\mathbf{I}= \left [
\begin{matrix}
  \frac{K}{\sigma_{v}} &  \frac{H}{\sigma_{v}}  & 0\\
  -\frac{H}{\sigma_{v}} & \frac{K}{\sigma_{v}} & 0\\
  0 &  0 & \frac{L}{\sigma_{v}}\\
\end{matrix}
\right ]
\end{equation}
and
\begin{equation}
\mathbf{J}= \left [
\begin{matrix}
  \sqrt{\sigma^2_{\perp}-\frac{K^2}{\sigma^2_{v}}-\frac{H^2}{\sigma^2_{v}}} &  0  & 0\\
  0 & \sqrt{\sigma^2_{\perp}-\frac{K^2}{\sigma^2_{v}}-\frac{H^2}{\sigma^2_{v}}} & 0\\
  0 &  0 & \sqrt{\sigma^2_{\parallel}-\frac{L^2}{\sigma^2_{v}}}\\
\end{matrix}
\right ] .
\end{equation}
The unit normal vectors $\mathbf{N_{1}}(0,1)$ and $\mathbf{N_{2}}(0,1)$, besides each having statistically independent Cartesian
components, are also by design statistically independent from each other, and may be generated by the Box-Muller method
\cite{Box1958} on a computer.

As expected, when $B=0$, $\mathbf{I}$ and $\mathbf{J}$ become diagonal matrices with their diagonal elements being
$L/\sigma_{v}$ and $\sqrt{\sigma^2_{\parallel}-L^2/\sigma^2_{v}}$, respectively, in which case Eqs.\ (\ref{equpdate}) become
\begin{eqnarray}
\mathbf{v}_{B=0}(t) & = & \langle\mathbf{v}\rangle + \sigma_{v}\mathbf{N_{1}}(0,1), \nonumber \\
\mathbf{r}_{B=0}(t) & = & \langle\mathbf{r}\rangle +  \frac{L}{\sigma_{v}} \mathbf{N_{1}}(0,1) +
\sqrt{\sigma^2_{\parallel}-\frac{L^2}{\sigma^2_{v}}}\mathbf{N_{2}}(0,1). \label{equpdate_B0}
\end{eqnarray}

It is worthwhile to point out that the updating formulae, Eqs.\ (\ref{equpdate}) and (\ref{equpdate_B0}), can be actually
interpreted as a two-step algorithm: first, one calculates the means of the velocity and position at time $t$, and secondly, one
adds explicitly random displacements (ERDs) of velocity and position. This is very important in the sense that the first step is
nothing more than solving deterministic Newton's equations with damping, so in principle many algorithms suitable for the MD
simulation (for example, the Verlet, Beeman and even multi-step PC algorithms \cite{Allen1989,Berendsen1986}) can be used here.
On the other hand, the second step is independent of the first one, and can always be performed at the end of the time step as a
correction to the previous step.

However, Eqs.\ (\ref{eqglmean}) and (\ref{eqglmean_B0}) are still not very suitable for computer simulations, because they
involve high-order derivatives of acceleration, which are not directly accessible in simulations. Further assumptions must be
made about $\mathbf{a}(t)$, and different ways of handling that issue then result in different simulation methods, such as the
Euler-like, Beeman-like, Verlet-like and Gear-like methods.

\section{Euler-like method}\label{Sec_EL}
In the Euler-like method \cite{Ermak}, it is assumed that $\mathbf{F}=\mathbf{F(0)}$, or $\mathbf{a}=\mathbf{a(0)}$ = constant
in the time interval $[0, t]$. Under this assumption, the velocity and position means are, respectively, given by
\begin{eqnarray}
\langle\mathbf{r}\rangle_{EL}&=&\mathbf{r}_{0} + \mathbf{I}_{1}\cdot\mathbf{v}_{0}t +
\mathbf{I}_{2}\cdot\mathbf{a}_{0}t^2, \nonumber \\
\langle\mathbf{v}\rangle_{EL}&=&\mathbf{I}_{0}\cdot\mathbf{v}_{0}+ \mathbf{I}_{1}\cdot\mathbf{a}_{0}t , \label{eqeuler}
\end{eqnarray}
where the subscript ${EL}$ in Eq.\ (\ref{eqeuler}) is added to indicate reference to the Euler-like method. The special case of
zero magnetic field can be simplified to \cite{Ermak}
\begin{eqnarray}
\langle\mathbf{r}\rangle_{EL,B=0}&=&\mathbf{r}_{0} + d_{1} \mathbf{v}_{0}t +
d_{2} \mathbf{a}_{0}t^2, \nonumber \\
\langle\mathbf{v}\rangle_{EL,B=0}&=&d_{0} \mathbf{v}_{0}+ d_{1} \mathbf{a}_{0}t. \label{eqeuler_B0}
\end{eqnarray}

The procedure for implementing the simulation by using Euler-like method is as follows: \\
(a) Evaluate Eq.\ (\ref{eqeuler}) [or Eq.\ (\ref{eqeuler_B0}) in the case of zero magnetic field] for the means of velocity and
displacement, given the initial conditions $\mathbf{v}_{0}$,
$\mathbf{r}_{0}$ and $\mathbf{a}_{0}$ at $t=0$. \\
(b) Generate two independent random vectors, $\mathbf{N_{1}}(0,1)$ and $\mathbf{N_{2}}(0,1)$, which follow the standard normal
distribution (this may be usually done by using the Box-Muller method \cite{Box1958}),
and substitute them into Eq.\ (\ref{equpdate}) [or Eq.\ (\ref{equpdate_B0}) in the case of zero magnetic field], along with the means obtained in the first step. \\
(c) After the above two steps, a force evaluation is done, which will provide an initial condition for $\mathbf{a}_{0}$, along
with the updated position and velocity, to be used in the next step.

\section{The Beeman-like method}\label{Sec_BL}
In the Beeman-like method \cite{Allen,Allen1989}, one has
\begin{eqnarray}
\langle\mathbf{r}\rangle_{BL}&=&\mathbf{r}_{0} + \mathbf{I}_{1}\cdot\mathbf{v}_{0}t +
\mathbf{I}_{2}\cdot\mathbf{a}_{0}t^2 + \mathbf{I}_{3}\cdot\mathbf{\dot{a}}_{0}t^3, \nonumber \\
\langle\mathbf{v}\rangle_{BL}&=&\mathbf{I}_{0}\cdot\mathbf{v}_{0}+ \mathbf{I}_{1}\cdot\mathbf{a}_{0}t +
\mathbf{I}_{2}\cdot\mathbf{\dot{a}}_{0}t^2 + \mathbf{I}_{3}\cdot \mathbf{\ddot{a}}_{0}t^3. \label{eqblmean}
\end{eqnarray}
Note that, according to the Beeman algorithm \cite{Allen,Allen1989}, terms through first order in $t$, i.e., $\mathbf{a}(t)
=\mathbf{a}_{0}+\mathbf{\dot{a}}_{0}t$, are kept in the expression for $\langle\mathbf{r}\rangle_{BL}$, and terms through second
order in $t$, i.e., $\mathbf{a}(t)=\mathbf{a}_{0}+\mathbf{\dot{a}}_{0}t +\frac{1}{2!}\mathbf{\ddot{a}}_{0}t^2$, are kept in
$\langle\mathbf{v}\rangle_{BL}$.

However, the above expressions are not yet suitable for a one-step simulation method \cite{Berendsen1986}, as they contain
derivatives of the acceleration, which have to be replaced by a finite difference formula in terms of $\mathbf{a}(0)$,
$\mathbf{a}(-t)$ and $\mathbf{a}(t)$. Thus, we obtain the schemes of the Beeman-like algorithm \cite{Allen,Allen1989},
\begin{eqnarray}
\langle\mathbf{r}\rangle_{BL}&=&\mathbf{r}_{0} + \mathbf{I}_{a}\cdot\mathbf{v}_{0}t +  \mathbf{I}_{b}\cdot\mathbf{a}_{0}t^2 +
 \mathbf{I}_{c}\cdot\mathbf{a}_{-t}t^2, \label{eqbeemanr} \\
\langle\mathbf{v}\rangle_{BL}&=& \mathbf{I}_{d}\cdot\mathbf{v}_{0}+
 \mathbf{I}_{e}\cdot\mathbf{a}_{t}t +  \mathbf{I}_{f}\cdot\mathbf{a}_{0}t +  \mathbf{I}_{g}\cdot \mathbf{a}_{-t}t,
\label{eqbeemanv}
\end{eqnarray}
where $\mathbf{a}_{t} \equiv \mathbf{a}(t)$, $\mathbf{a}_{-t} \equiv \mathbf{a}(-t)$, and
\begin{align*}
\mathbf{I}_{a}&=\mathbf{I}_{1}, &\mathbf{I}_{b}&=\mathbf{I}_{2}+\mathbf{I}_{3},
&\mathbf{I}_{c}&=-\mathbf{I}_{3} \nonumber \\
\mathbf{I}_{d}&=\mathbf{I}_{0}, &\mathbf{I}_{e}&=\mathbf{I}_{2}-\mathbf{I}_{0}\mathbf{I}_{3}/\mathbf{I}_{1},
&\mathbf{I}_{f}&=\mathbf{I}_{1}-\mathbf{I}_{2}+2\mathbf{I}_{0}\mathbf{I}_{3}/\mathbf{I}_{1},
&\mathbf{I}_{g}&=-\mathbf{I}_{0}\mathbf{I}_{3}/\mathbf{I}_{1}. \label{eqcbeeman}
\end{align*}
[Note that, here, matrix operations should be understood as direct element operations, e.g.,
$(\mathbf{I}_{0}\mathbf{I}_{3})_{ij}=\mathbf{I}_{0,ij}\mathbf{I}_{3,ij}$.]

In the special case of zero magnetic field, Eqs. (\ref{eqbeemanv}) and (\ref{eqbeemanr}) can be simplified to \cite{Allen}
\begin{eqnarray}
\langle\mathbf{r}\rangle_{BL,B=0}&=&\mathbf{r}(0) + c_{a}\mathbf{v}(0)t + c_{b}\mathbf{a}(0)t^2 +
c_{c}\mathbf{a}(-t)t^2, \label{eqbeemanr_B0} \\
\langle\mathbf{v}\rangle_{BL,B=0}&=&c_{d}\mathbf{v}(0)+ c_{e}\mathbf{a}(t)t + c_{f}\mathbf{a}(0)t + c_{g} \mathbf{a}(-t)t,
\label{eqbeemanv_B0}
\end{eqnarray}
with
\begin{align*}
c_{a}&=d_{1}, &c_{b}&=d_{2}+d_{3}, &c_{c}&=-d_{3} \nonumber \\
c_{d}&=d_{0}, &c_{e}&=d_{2}-d_{0}d_{3}/d_{1}, &c_{f}&=d_{1}-d_{2}+2d_{0}d_{3}/d_{1}, &c_{g}&=-d_{0}d_{3}/d_{1}.
\end{align*}

Implementation of the Beeman-like method is also based on Eq.\ (\ref{equpdate}), and the detailed simulation procedure
is listed in the following:\\
(a) Evaluate Eq.\ (\ref{eqbeemanr}) [or Eq.(\ref{eqbeemanr_B0})] for the mean of displacement, given the initial conditions
$\mathbf{v}_{0}$,
$\mathbf{r}_{0}$ and $\mathbf{a}_{0}$ at $t=0$, and $\mathbf{a}(-t)$ at time $-t$. \\
(b) Evaluate a new deterministic acceleration $\mathbf{a}(t)$ based on the new position obtained in the previous step.\\
(c) Evaluate Eq.\ (\ref{eqbeemanv}) [or Eq.(\ref{eqbeemanv_B0})] for the mean of velocity, given the initial conditions
$\mathbf{v}_{0}$ and $\mathbf{a}_{0}$ at $t=0$, and $\mathbf{a}(-t)$ at time $-t$, as well as the newly obtained acceleration
$\mathbf{a}(t)$ in the previous step. [Note that, if one uses a periodic boundary condition,
it should be applied before this step.]\\
(d) Generate two independent random vectors, $\mathbf{N_{1}}(0,1)$ and $\mathbf{N_{2}}(0,1)$, which follow the standard normal
distribution, and substitute them into Eq.\ (\ref{equpdate}) along with the means
obtained in the steps (a) and (c).\\
(e) Apply periodic boundary condition again if using it.

As one can see, the difference compared to the Euler-like method, is that here one has to do the force evaluation after the
position update, but before the velocity update, in every time step, and one has to store the force value of the last step to be
used for $\mathbf{a}(-t)$. In addition, one needs to apply the periodic boundary condition twice.

Let us note that the Verlet-like method \cite{vanGunsteren1982} was based on similar assumptions for $\mathbf{a}(t)$ as the
Beeman-like method, and it was proven \cite{Allen} that the Verlet-like method is numerically equivalent to the Beeman-like
method in the position, while the latter seems to have a better accuracy in the velocity. Therefore, extensions of the
Verlet-like method are not presented here.

\section{Gear-like Predictor-Corrector method}\label{Sec_GL}
A Gear-like Predictor-Corrector (PC) method for BD \cite{Hou2008pop} can be constructed in a direct analogy with the Gear method
for MD simulations. Our Gear-like method also includes three stages, namely, predicting, force evaluating, and correcting
\cite{Allen1989,Berendsen1986}, as in the MD simulation, but the difference here is that one has to add explicit random
displacements of velocity and position by using Eq.\ (\ref{equpdate}) at the end of time step to complete the BD simulation. The
basic procedure goes as follows.

\subsection{Predicting}
In the predicting stage, one has
\begin{eqnarray}
\langle\mathbf{r}\rangle^{P}&=&\mathbf{r}_{0} + \mathbf{I}_{1}\cdot\mathbf{v}_{0}t + \mathbf{I}_{2}\cdot\mathbf{a}_{0}t^2 +
\mathbf{I}_{3}\cdot\mathbf{\dot{a}}_{0}t^3 +
\mathbf{I}_{4}\cdot\mathbf{\ddot{a}}_{0}t^4 + \mathbf{I}_{5}\cdot\mathbf{\dddot{a}}_{0}t^5 ,\nonumber \\
\langle\mathbf{v}\rangle^{P}&=&\mathbf{I}_{0}\cdot\mathbf{v}_{0}+ \mathbf{I}_{1}\cdot\mathbf{a}_{0}t +
\mathbf{I}_{2}\cdot\mathbf{\dot{a}}_{0}t^2 + \mathbf{I}_{3}\cdot \mathbf{\ddot{a}}_{0}t^3 +
\mathbf{I}_{4}\cdot\mathbf{\dddot{a}}_{0}t^4,  \nonumber \\
\mathbf{a}^{P} &=& \mathbf{a}_{0} + \dot{\mathbf{a}}_{0}t + \frac{1}{2!}\ddot{\mathbf{a}}_{0}t^2 +
\frac{1}{3!}\dddot{\mathbf{a}}_{0}t^3, \nonumber \\
\mathbf{\dot{a}}^{P} &=& \mathbf{\dot{a}}_{0} +
\mathbf{\ddot{a}}_{0}t + \frac{1}{2!}\mathbf{\dddot{a}}_{0}t^2, \nonumber \\
\ddot{\mathbf{a}}^{P} &=&
\mathbf{\ddot{a}}_{0} + \mathbf{\dddot{a}}_{0}t, \nonumber \\
\dddot{\mathbf{a}}^{P} &=& \mathbf{\dddot{a}}_{0}, \label{eqpredictor}
\end{eqnarray}
where the superscript $P$ indicates that these are quantities in the predicting stage. For simplicity, we have dropped in the
above all derivatives of $\mathbf{a}(t)$ higher than the third order, but we note that extensions to higher orders are quite
straightforward. One notices in Eq.\ (\ref{eqpredictor}) that we have used Eq.\ (\ref{eqglmean}) for the means of position and
velocity, instead of using Taylor series for position and velocity which usually provide a basis for implementing the Gear
method in the MD simulation \cite{Berendsen1986,Allen1989}. In the case of zero magnetic field, one simply needs to replace the
means of velocity and position given by Eq.\ (\ref{eqglmean_B0}), instead of  Eq.\ (\ref{eqglmean}). The rest of the algorithm
(derivatives of the force) is essentially the same as in the MD.

\subsection{Force evaluating}
In the next step, the predicted position $\mathbf{r}^{P}(t)$ is used to obtain a new force, or deterministic acceleration
$\mathbf{a}(t)$, and a difference between the predicted acceleration $\mathbf{a}^{P}(t)$ and the new acceleration
$\mathbf{a}(t)$ is calculated as
\begin{equation}
\Delta\mathbf{a}\equiv\mathbf{a}(t)- \mathbf{a}^{P}(t). \label{eqfe}
\end{equation}
It can be seen that this step is exactly the same as the one normally used for the Gear method in MD
\cite{Allen1989,Berendsen1986}.

\subsection{Correcting}
In the correcting stage, the above difference term is further used to correct all predicted positions and their "derivatives",
thus giving
\begin{eqnarray}
\langle\mathbf{r}\rangle^{C} &=& \langle\mathbf{r}\rangle^{P} + 2\alpha_{0}\mathbf{I}_{2}\cdot\Delta \mathbf{R}, \nonumber
\\
\langle\mathbf{v}\rangle^{C}t &=& \langle\mathbf{v}\rangle^{P}t + \alpha_{1}\mathbf{I}_{1}\cdot\Delta \mathbf{R}, \nonumber
\\
\frac{\mathbf{a}^{C}t^2}{2!} &=& \frac{\mathbf{a}^{P}t^2}{2!} + \alpha_{2}\Delta \mathbf{R}, \nonumber
\\
\frac{\mathbf{\dot{a}}^{C}t^3}{3!} &=& \frac{\mathbf{\dot{a}}^{P} t^3}{3!} + \alpha_{3}\Delta \mathbf{R}, \nonumber
\\
\frac{\mathbf{\ddot{a}}^{C}t^4}{4!} &=& \frac{\mathbf{\ddot{a}}^{P}t^4}{4!} + \alpha_{4}\Delta \mathbf{R}, \nonumber
\\
\frac{\mathbf{\dddot{a}}^{C}t^5}{5!} &=& \frac{\mathbf{\dddot{a}}^{P}t^5}{5!} + \alpha_{5}\Delta \mathbf{R}, \label{eqcorrector}
\end{eqnarray}
where
\begin{equation}
\Delta \mathbf{R} \equiv \frac{\Delta\mathbf{a}t^2}{2!}, \label{eqdelta}
\end{equation}
and the coefficients $\alpha_i$ are given in the Table \ref{table1}. Note that the table is simply a reproduction of those
appearing in Refs. \cite{Allen1989,Berendsen1986}, and is given here for completeness. By using parameters in different columns
of that table, one can achieve 3rd-, 4th-, and 5th-order (or 4-, 5- and 6-value) \cite{Allen1989,Berendsen1986} Gear-like
algorithms for the BD simulation. Note that the first two equations in Eq.\ (\ref{eqcorrector}) are slightly different from
those in the MD in order to restore the damping effect on deterministic acceleration, and to maintain consistence with the
corresponding terms in Eq.\ (\ref{eqglmean}) [or the first two equations in Eq.\ (\ref{eqpredictor})] as well. In the case of
zero magnetic field, one simply replaces $\mathbf{I}_{0}$ and $\mathbf{I}_{1}$ in Eq. (\ref{eqcorrector}) by $d_{0}$ and
$d_{1}$, respectively.

\begin{center}
\begin{table}
\caption{Coefficients used in correcting stage of Gear-like PC method. Note that the table is simply a reproduction of those
appearing in Refs. \cite{Allen1989,Berendsen1986}.}
\begin{tabular*}{0.75\textwidth}{@{\extracolsep{\fill}}cccc}\hline\hline
$\alpha_{i}$ &      3th-order      &      4th-order           &     5th-order \\
\hline
$\alpha_{0}$ & $1/6$ & $19/120$   & $3/16$ \\
$\alpha_{1}$ & $5/6$ & $3/4$      & $251/360$ \\
$\alpha_{2}$ &      1        &         1          &      1           \\
$\alpha_{3}$ & $1/3$ & $1/2$      & $11/18$ \\
$\alpha_{4}$ &       0       & $1/12$     & $1/6$ \\
$\alpha_{5}$ &       0       &         0          & $1/60$ \\
\hline\hline
\end{tabular*} \label{table1}
\end{table}
\end{center}

\subsection{Adding explicitly random displacements (ERDs)}
To complete the BD simulation, we have to use the updating formula, Eq.\ (\ref{equpdate}), to add ERDs of the velocity and
position. It should be noted that now the corrected value $\langle\mathbf{r}\rangle^{C}$ and $\langle\mathbf{v}\rangle^{C}$ must
be used, respectively, in the places of $\langle\mathbf{r}\rangle$ and $\langle\mathbf{v}\rangle$ in Eq.\ (\ref{equpdate}).

We summarize the basic steps for implementing the Gear-like PC method:\\
(a) Eq.\ (\ref{eqpredictor}) is used to calculate predicted values of the position, velocity, acceleration and its derivatives,
given the initial conditions, $\mathbf{r}_{0}$, $\mathbf{v}_{0}$, $\mathbf{a}_{0}$, $\mathbf{\dot{a}}_{0}$,
$\mathbf{\ddot{a}}_{0}$ and $\mathbf{\dddot{a}}_{0}$, at $t=0$. Note that, for the very first few steps of the simulation,
$\mathbf{\dot{a}}_{0}$, $\mathbf{\ddot{a}}_{0}$ and $\mathbf{\dddot{a}}_{0}$ are undefined. The simplest way to get around this
issue is to simply set all of them to zero at the very first step, and their values then will be updated during subsequent
iterations. A better way would be to start the simulation by using a Runge-Kutta procedure for the first few steps
\cite{Berendsen1986}. However, neither of these alternatives will have any significant
effects on the results in real many-particle simulations.\\
(b) Evaluate new acceleration $\mathbf{a}(t)$ by using the predicted position $\langle\mathbf{r}\rangle^P$, and calculate its
difference with the predicted value $\mathbf{a}^{P}(t)$ by using Eq.\ (\ref{eqfe}).
[Note that, if one uses periodic boundary condition, it should be applied before the force evaluation.]\\
(c) Correct the predicted values of the position, velocity, acceleration and its derivatives by using
Eq.\ (\ref{eqcorrector}).\\
(d) Generate two independent random vectors, $\mathbf{N_{1}}(0,1)$ and $\mathbf{N_{2}}(0,1)$, which follow the standard normal
distribution, and substitute them into Eq.\ (\ref{equpdate}) together with the corrected values
$\langle\mathbf{r}\rangle^{C}$ and $\langle\mathbf{v}\rangle^{C}$.\\
(e) Apply periodic boundary condition again if using it.

The above is a basic procedure for using the Gear-like PC method for the BD. One might have noticed that the formulas, as well
as the simulation procedure, are quite similar to those used in the Gear method for the MD based on Newton's equations
\cite{Berendsen1986,Allen1989}, apart from our use of Eqs.\ (\ref{eqpredictor}) and (\ref{eqdelta}) to express the velocity and
position, as well as the addition of ERDs at the end of every time step. Also, when $B=0$ and $\gamma\rightarrow 0$, the
Gear-like method goes over to the Gear method for MD simulation.

\section{Testing the algorithms}
In the above we have developed numerical algorithms for simulating Brownian dynamics of charged particles in an external
magnetic field. But how accurate are they in describing the actual Brownian motion? To answer this question, we present in this
section some simple computational examples as testing cases and compare the performances of different algorithms presented
above. For simplicity, we shall occasionally denote the Euler-like, Beeman-like, Verlet-like and Gear-like methods by EL, BL,
VL, and GL, respectively.

As the simplest test cases, one could adopt comparison of the numerical results with the analytical results for some simple
model problems, such as the classical harmonic oscillator, which had been used extensively to test algorithms for the MD
simulation (see for example Ref.\ \cite{Berendsen1986} and \cite{Venneri1987} for nice reviews). We follow here the same logic and employ the model of a
three-dimensional (3D) Brownian-harmonic-oscillator (BHO) in an external magnetic field \cite{Jimenez2008}, for which the
Langevin equation (\ref{eqlangevin}) is reduced to
\begin{eqnarray}
\frac{dv_x}{dt}&=&-\gamma v_{x} - \omega_{0}^{2}x + \Omega v_{y} + A_{x}(t) ,\nonumber \\
\frac{dv_y}{dt}&=&-\gamma v_{y} - \omega_{0}^{2}y - \Omega v_{x} + A_{y}(t) ,\nonumber \\
\frac{dv_z}{dt}&=&-\gamma v_{z} - \omega_{0}^{2}z +                A_{z}(t) . \label{eqsdho0}
\end{eqnarray}
Since the magnetic field is in the $z$ direction of the Cartesian coordinate system, two independent processes take place. In
the $z$ direction, the magnetic field does not have any effect on Brownian motion, so the Brownian particle behaves like a
one-dimensional stochastically-damped harmonic-oscillator \cite{Chandrasekhar1943,Lemons2002}, while in the directions
perpendicular to the magnetic field, i.e., in the $xy$ plane, Brownian motion is much more complicated because of the coupling
between the motions in the $x$ and $y$ directions via magnetic field. It is curious to note that, although the analytical
solution for a BHO without magnetic field had been known for well over a half of the century \cite{Chandrasekhar1943}, the
problem of BHO in an external magnetic field has been solved analytically only very recently by Jim\'{e}nez-Aquino \textit{et
al.}\cite{Jimenez2008}.

\begin{figure}
\centering
\includegraphics[trim=10mm 10mm 10mm 10mm, width=0.8\textwidth]{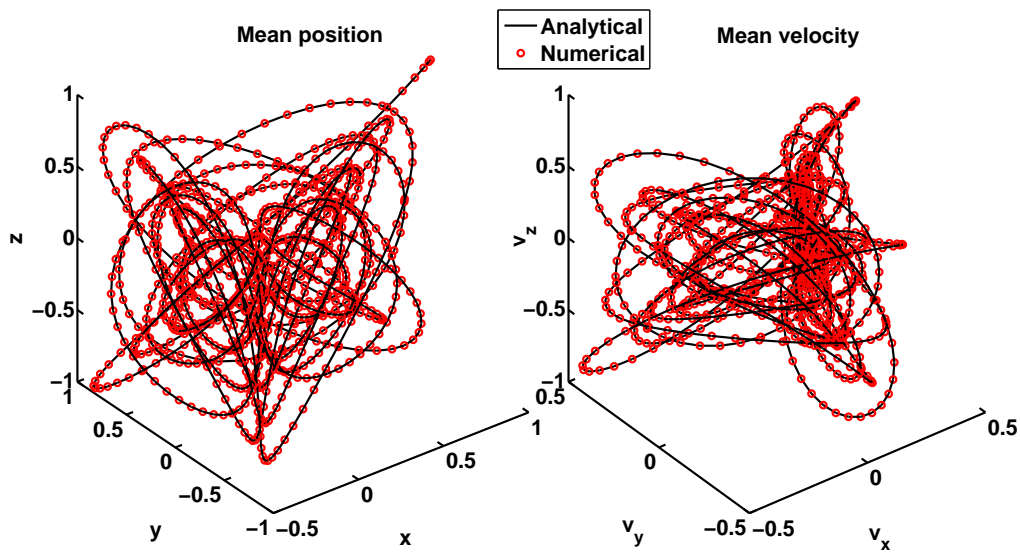}
\caption{(Color online) Mean position (left) and mean velocity (right) of a Brownian oscillator in magnetic field during $200$
time units, with $\gamma=0.02$, $\omega_{0}=1/\sqrt{2}$, $\Omega=0.5$ and $\Delta t=0.02$. Initial conditions are $x_{0}=1.0$,
$y_{0}=0$, $z_{0}=1.0$ and $v_{x0}=v_{y0}=v_{z0}=0$. Solid lines are the result of analytical solutions \cite{Jimenez2008}, and
circles are the numerical results calculated by using the GL-5 method.} \label{fig_mean_xyz}
\end{figure}

\begin{figure}
\centering
\includegraphics[trim=10mm 10mm 10mm 10mm, width=0.8\textwidth]{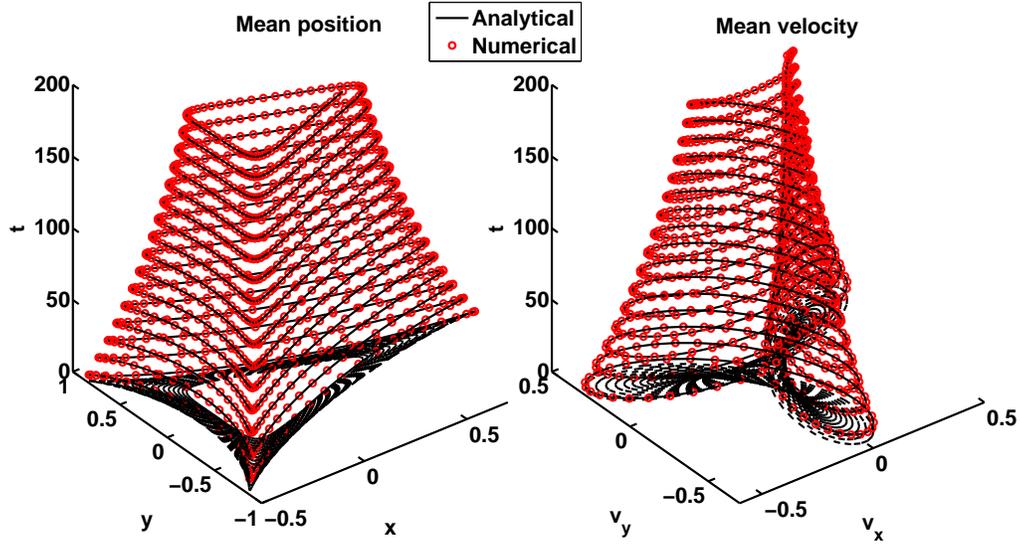}
\caption{(Color online) Components of the mean position (left) and mean velocity (right) in the $xy$ plane of a Brownian
oscillator in magnetic field during $200$ time units, with $\gamma=0.02$, $\omega_{0}=1/\sqrt{2}$, $\Omega=0.5$ and $\Delta
t=0.02$. Initial conditions are the same as in fig.\ \ref{fig_mean_xyz}. Solid lines are the result of analytical solutions
\cite{Jimenez2008}, and circles are the numerical results calculated by using the GL-5 method. Dashed lines are projections of
the analytical results on the $xy$ plane.} \label{fig_mean_xyt}
\end{figure}

Of course, the dynamics a BHO in magnetic field can be also traced by using the above described numerical methods. It is
important to realize that, although the physical model for a BHO in magnetic field, Eq.\ (\ref{eqsdho0}), involves a very simple
deterministic acceleration which depends on the particle's position, it nevertheless provides a good test for our assumption
that this acceleration depends on time only, and to examine the effects of various truncation schemes for the Taylor series
representation of this acceleration, i. e., Eq.\ (\ref{eqtaylor}). The results of our
numerical simulations of the BHO in a magnetic filed will be compared with the corresponding explicit analytical solutions, for
which we refer the reader to the original reference \cite{Jimenez2008}. By doing so, one can evaluate the accuracy and
performance of the numerical methods and validate the assumptions made in their derivation. The results will serve as a basic
reference for future simulations of more complicated systems, such as magnetized dusty plasmas \cite{Venneri1987}.

According to the above discussion, and particularly referring to Eq.\ (\ref{equpdate}), the task of a BD simulation is simply to
predict the position $\mathbf{r}(t)$ and velocity $\mathbf{v}(t)$ of a Brownian particle at time $t$, given the set of initial
conditions at time $0$. Since $\mathbf{r}(t)$ and $\mathbf{v}(t)$ can be obtained numerically in two-steps, first, by
calculating the means of the velocity and position at time $t$ and, second, by adding the ERDs of velocity and position, the
performance of a BD simulation will be examined by testing the accuracy of both steps. We begin by testing the first step, i.e.,
calculating the means.

\subsection{Inaccuracy in calculating the means}
\begin{figure}
\centering
\includegraphics[trim=0 10mm 150mm 0mm, width=0.7\textwidth]{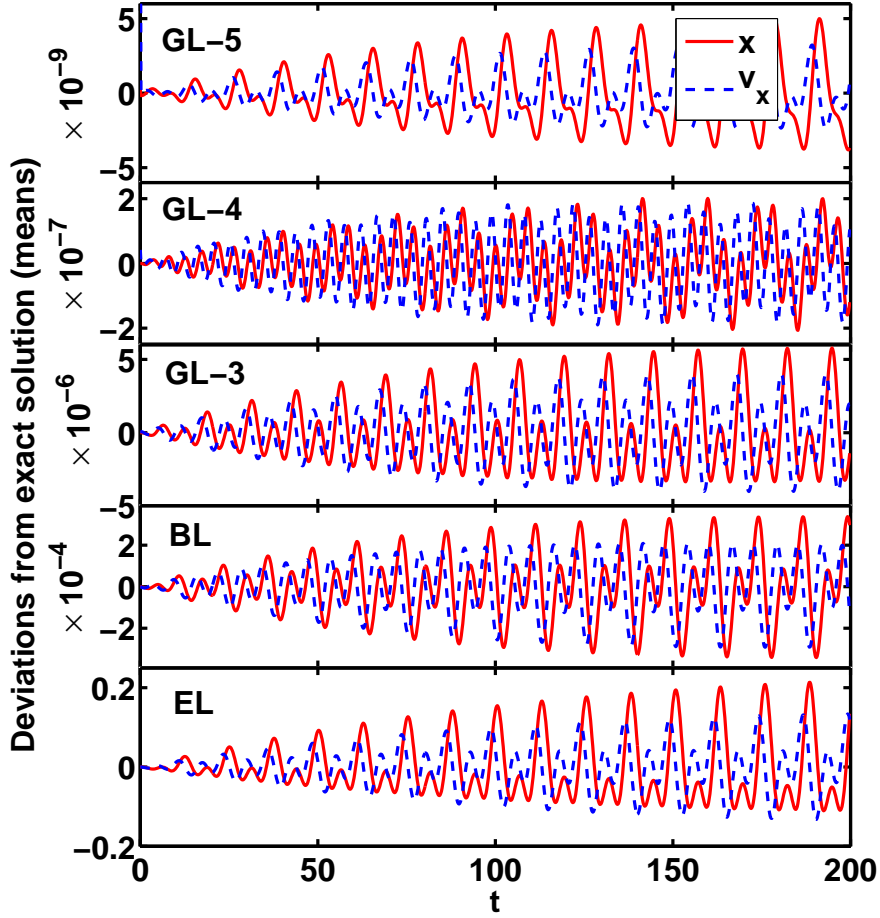}
\caption{(Color online) Deviations of the mean $x$ (solid lines) and mean $v_x$ (dashed lines) values from the analytical
solutions \cite{Jimenez2008} using different methods (from top to bottom: GL-5, GL-4, GL-3, BL and EL), with $\gamma=0.02$,
$\omega_{0}=1/\sqrt{2}$, $\Omega=0.5$ and $\Delta t=0.02$. Initial conditions are the same as in Fig.\ \ref{fig_mean_xyz}. Note
the magnitudes of the errors.} \label{fig_dev_mean_x}
\end{figure}

Without any loss of generality, here and in the subsequent simulations we set $k_{B}T=1$, $m=1$ and $\omega_{0}=\sqrt{2}/2$, and
choose the initial conditions to be $x_{0}=1.0$, $y_{0}=0$, $z_{0}=1.0$ and $v_{x0}=v_{y0}=v_{z0}=0$.

We first present in Fig.\ \ref{fig_mean_xyz} examples of full 3D trajectories of the mean position and velocity for a BHO in
$200$ time units, with $\gamma=0.02$, $\Omega=0.5$ and the time step size $\Delta t=0.02$. Solid lines are the results of the
analytical solutions [Eqs.\ (B23)-(B34) in \cite{Jimenez2008}], while the circles are numerical results calculated by using the
GL-5 method. One might be particularly interested in the motion in the directions perpendicular to the magnetic field, which is
shown in Fig.\ \ref{fig_mean_xyt} as the time evolutions of the mean position and the mean velocity in the $xy$ plane. We have
found that details in the trajectory patterns strongly depend on the initial conditions and on the values of $\omega_{0}$ and
$\Omega$, but the general tendency in trajectories is the same. Without Brownian acceleration, the initial energy of the
oscillator would be sooner or later consumed by the damping, and the oscillator would come to rest at $x=y=z=0$ after long
enough time. One sees from Figs.\ \ref{fig_mean_xyz} and \ref{fig_mean_xyt} that the numerical results agree very well with the
analytical solutions. Good visual agreements were also found with numerical results of the GL-4, GL-3 and BL. However,
quantitatively, they are quite different, as is shown in the following.

\begin{figure}
\centering
\includegraphics[trim=150mm 10mm 0 0mm, width=0.7\textwidth]{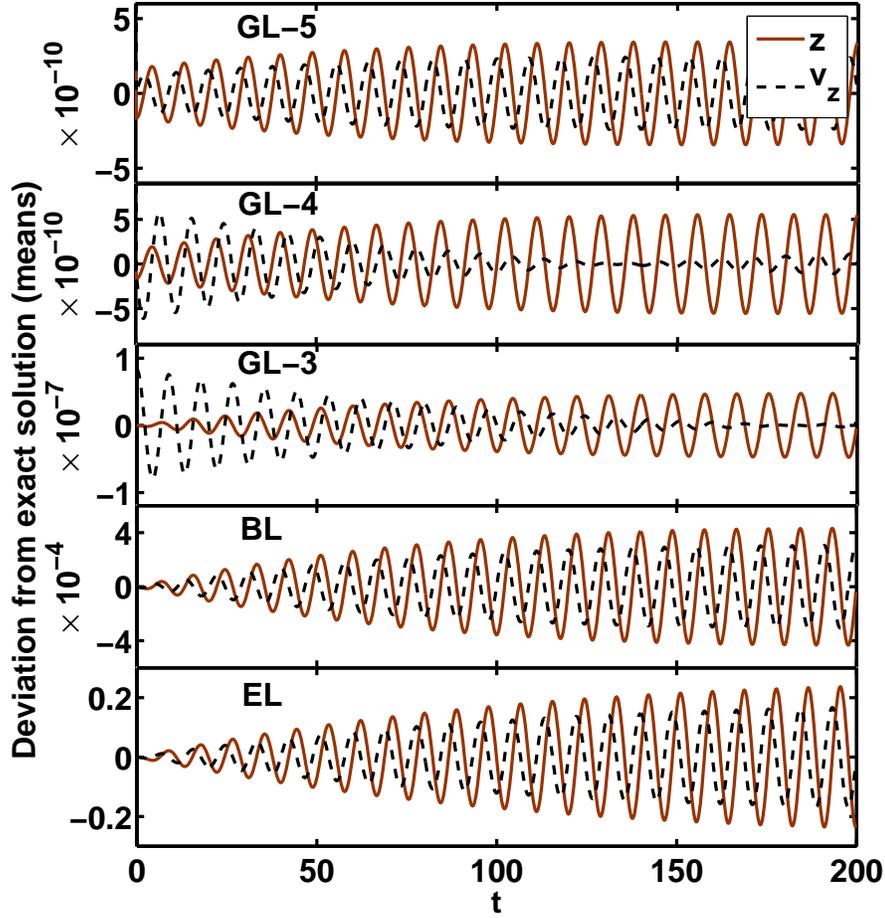}
\caption{(Color online)  Deviations of the mean $z$ (solid lines) and mean $v_z$ (dashed lines) from the analytical solutions
\cite{Jimenez2008} using different methods (from top to bottom: GL-5, GL-4, GL-3, BL and EL) in $200$ time units, with
$\gamma=0.02$, $\omega_{0}=1/\sqrt{2}$, $\Omega=0.5$ and $\Delta t=0.02$. Initial conditions are the same as in Fig.\
\ref{fig_mean_xyz}. Note the magnitude of the errors.} \label{fig_dev_mean_z}
\end{figure}

Figures \ref{fig_dev_mean_x} and \ref{fig_dev_mean_z} display deviations (differences) of the numerical results for the means of
the position and velocity from the corresponding analytical results \cite{Jimenez2008} in $200$ time units, under the same
conditions as in Fig.\ \ref{fig_mean_xyz}. Only deviations in the $x$ and $z$ directions are shown in Figs.\
\ref{fig_dev_mean_x} and \ref{fig_dev_mean_z}, respectively, as those in the $y$ direction are essentially the same as those for
$x$, apart from a phase shift. The oscillatory patterns of deviations in the position and velocity approximately resemble those
of the full solutions in Fig.\ \ref{fig_mean_xyz}, but they have much smaller amplitudes. One can observe the differences in
magnitude of deviations for different methods, with the GL-5 method having the smallest deviations and therefore highest
accuracy in both the $x$ and $z$ directions, while the EL method exhibits the largest deviations, as can be expected from our
previous comparisons of these methods. It should also be noted that the performance of these methods is different in the $x$ and
$z$ directions. One sees generally higher accuracy in the $z$ direction for the Gear-like methods, while the BL and EL methods
have similar accuracies in the two directions. This indicates that the presence of the magnetic field also affects the accuracy
of the computation.

\begin{figure}
\centering
\includegraphics[trim=80mm 10mm 50mm 10mm, width=0.7\textwidth]{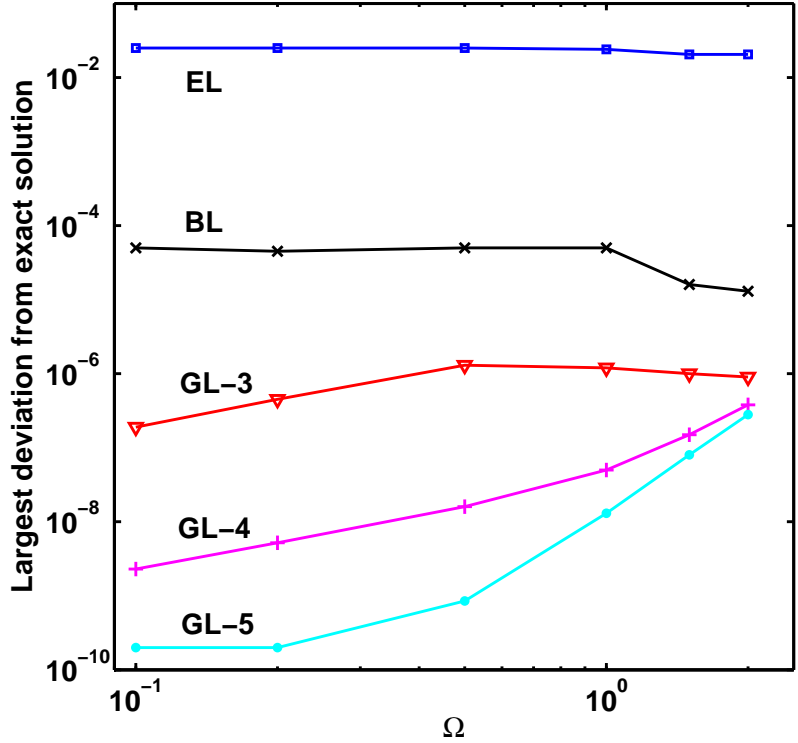}
\caption{(Color online) The largest deviation in the position ($x$-component) from the corresponding analytical solution
\cite{Jimenez2008} in the first $20$ time units versus $\Omega$, showing the dependence of inaccuracy of different methods on
the magnetic field, with $\omega_{0}=1/\sqrt{2}$, $\gamma=0.02$ and $\Delta t=0.02$.} \label{fig_dev_vs_Omega}
\end{figure}

A more detailed analysis of the dependence of the accuracy on the magnitude of magnetic field is shown in Fig.\
\ref{fig_dev_vs_Omega}, where the largest deviations of the position in the $x$ direction from the corresponding analytical
solution \cite{Jimenez2008} are recorded in the first $20$ time units (which is a convention for the measure of accuracy
\cite{Berendsen1986,Allen1989}) and are plotted versus $\Omega$ for different methods. One observes that, with the increase of
the magnetic field intensity, the largest deviation of the EL method remains almost constant, while that of the BL method
initially stays constant, but drops slightly when $\Omega>1.0$. The tendency observed in the Gear-like methods is the opposite.
The initially excellent accuracy deteriorates with increasing magnetic field. The one with the highest accuracy, i.e., the GL-5
method is affected the most, as is shown. A similar scaling rule applies to simulations with other time steps and damping rates,
as discussed next.

\begin{figure}
\centering
\includegraphics[trim=30 50mm 40 0mm, width=0.8\textwidth]{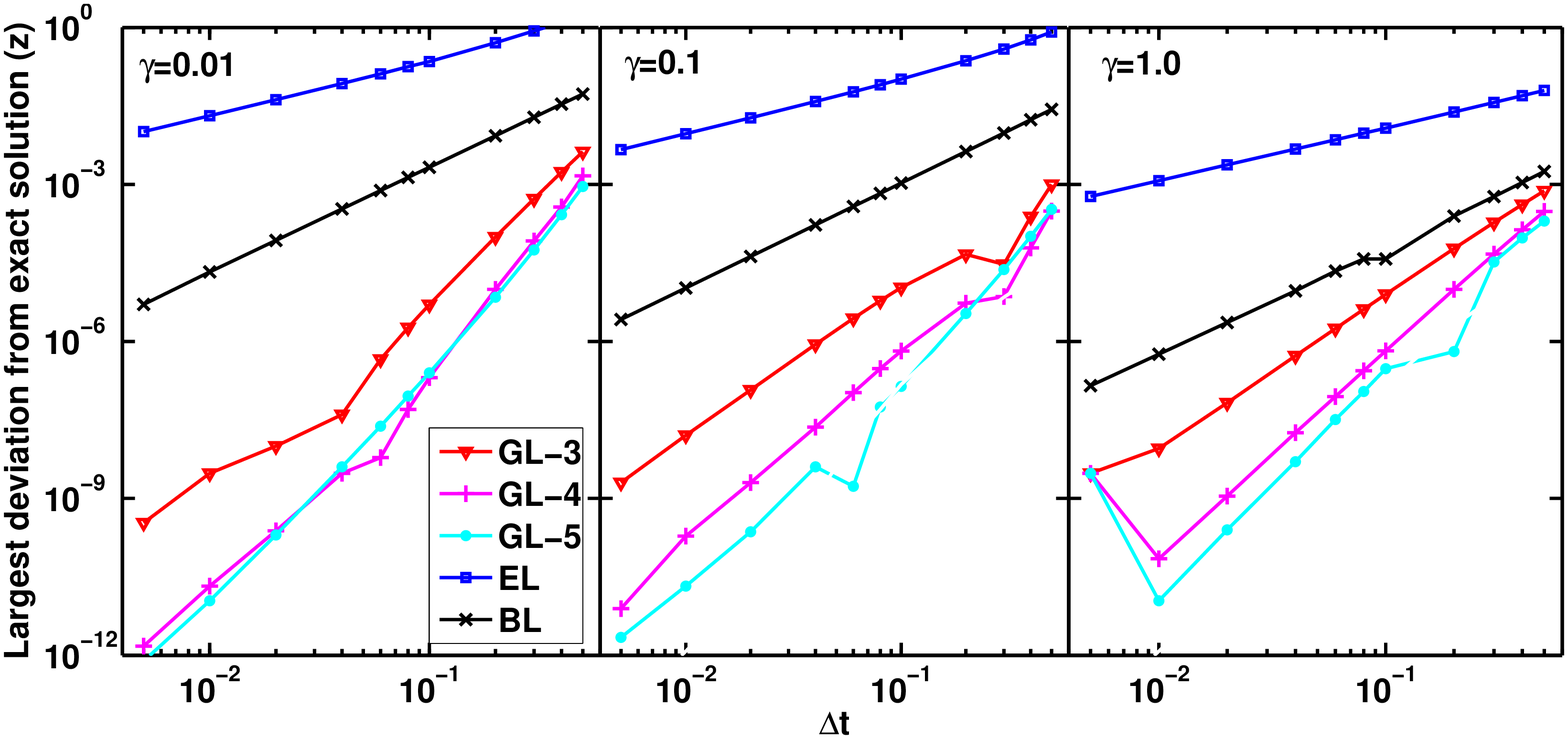}
\caption{(Color online) The largest deviation in the position ($z$-component) from the corresponding analytical solution
\cite{Jimenez2008} in the first $20$ time units versus time step size $\Delta t$, with $\omega_{0}=1/\sqrt{2}$ and $\Omega=0$,
for different methods and different $\gamma$ values.} \label{fig_dev_vs_dt}
\end{figure}

We make comparisons involving different time steps because that is always a key issue in both the MD and BD simulations. Fig.\
\ref{fig_dev_vs_dt} shows the largest deviations (defined by the largest deviation in the first 20 units) of the position
 in the $z$ direction versus the size $\Delta t$ of the time step for different methods and different friction coefficients $\gamma$.
This figure is plotted in a double logarithmic scale, and the curves are nearly straight lines. The slope of these lines is
called the \emph{apparent order} \cite{Berendsen1986}, illustrating the dependence of the error on the time step size. Namely,
if the error is found to be proportional to $\Delta t^{p}$, then the exponent $p$ is the \emph{apparent order}. It is found that
for very small $\gamma$, for example $\gamma=0.01$, as shown in Fig.\ \ref{fig_dev_vs_dt}, the apparent orders are
$p_{EL}\approx 1$, $p_{BL}\approx 2$, $p_{GL-3}\approx 3.5$, $p_{GL-4}\approx 4.2$ and $p_{GL-5}\approx 4.6$, respectively for
the Euler-like, Beeman-like, 3rd-, 4th-, and 5th-order Gear-like methods. These values are very close to those from the MD
simulations where damping is absent \cite{Berendsen1986}, which proves the consistency of our computation. When $\gamma$
increases, the absolute value of the error for Euler-like and Beeman-like methods decreases, while $p_{EL}$ and $p_{BL}$ remain
almost unchanged. On the other hand, $p_{GL-3}$, $p_{GL-4}$ and $p_{GL-5}$ slightly decrease with increasing $\gamma$. For
example, for $\gamma=1$, as shown in Fig.\ \ref{fig_dev_vs_dt}, the Gear-like methods have the worst performance in accuracy:
their apparent orders become now $p_{GL-3}\approx 3.1$, $p_{GL-4}\approx 3.9$ and $p_{GL-5}\approx 4.0$), but the magnitudes of
their errors are still much smaller than those of the Beeman-like and Euler-like methods.

All the above tests show that the numerical methods, especially the Gear-like methods and Beeman-like method, can describe the
mean values of the movement of a BHO with sufficiently high accuracy.

\begin{figure}
\centering
\includegraphics[trim=10 10mm 10 10mm, width=0.8\textwidth]{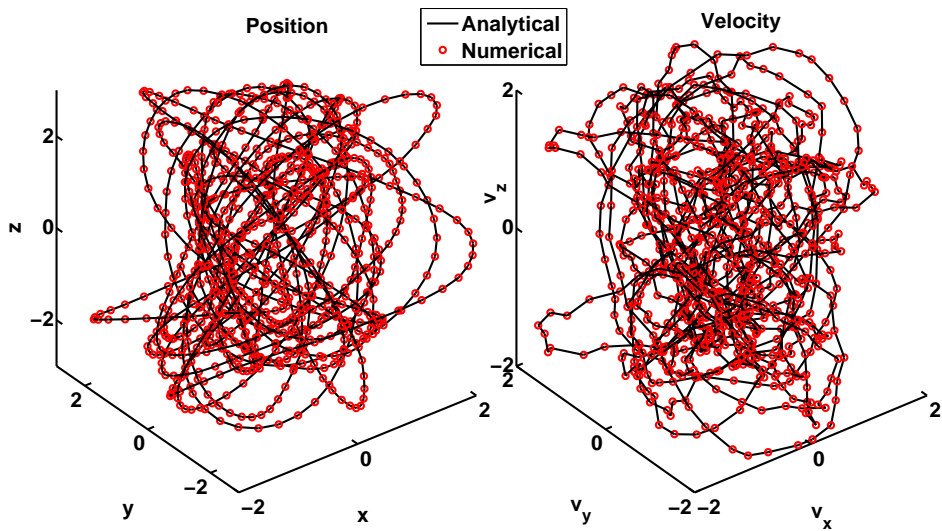}
\caption{(Color online)  Position (left) and velocity (right) of a Brownian oscillator in a magnetic field during $200$ time
units, with $\gamma=0.02$, $\omega_{0}=1/\sqrt{2}$, $\Omega=0.5$ and $\Delta t=0.02$. Initial conditions are the same as in
Fig.\ \ref{fig_mean_xyz}. Solid lines are the result of analytical solutions \cite{Jimenez2008}, and circles are the numerical
results calculated by using the GL-5 method.} \label{fig_all_xyz}
\end{figure}

\begin{figure}
\centering
\includegraphics[trim=10 10mm 10 10mm, width=0.8\textwidth]{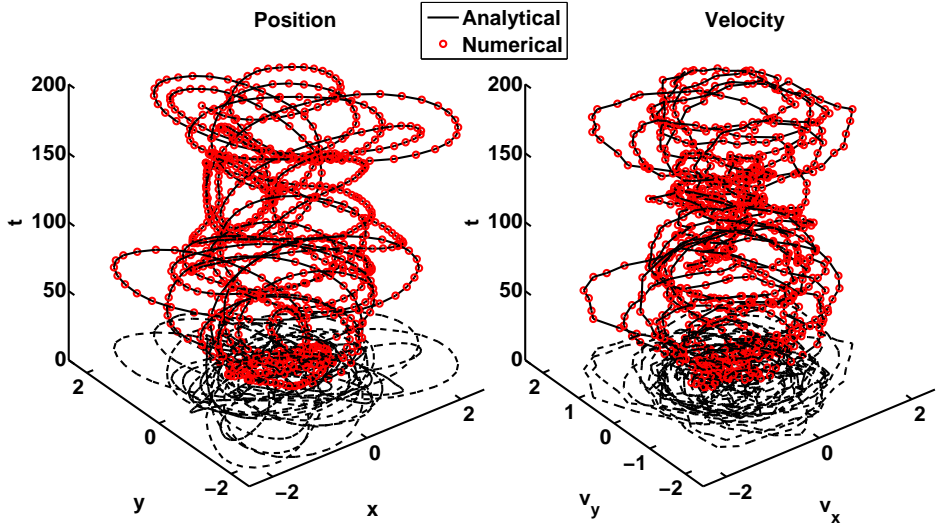}
\caption{(Color online) The $xy$ components of the position (left) and velocity (right) of a Brownian oscillator in a magnetic
field during $200$ time units, with $\gamma=0.02$, $\omega_{0}=1/\sqrt{2}$, $\Omega=0.5$ and $\Delta t=0.02$. Initial conditions
are the same as in Fig.\ \ref{fig_mean_xyz}. Solid lines are the result of analytical solutions \cite{Jimenez2008}, and circles
are the numerical results calculated by using the GL-5 method. Dashed lines are projections of the analytical results on the
$xy$ plane.} \label{fig_all_xyt}
\end{figure}

\subsection{Total inaccuracy}
The tests carried out in this sub-section are similar to those in the previous one, but with the addition of the ERDs in both
the position and velocity. We first show in Fig.\ \ref{fig_all_xyz} full trajectories of the BHO in $200$ time units, under the
same condition as in Fig.\ \ref{fig_mean_xyz}. Again, solid lines show results of the analytical solutions \cite{Jimenez2008},
while circles are the numerical results calculated by using the GL-5 method. Motion of the BHO in the $xy$ plane under the
influence of magnetic field is shown in Fig.\ \ref{fig_all_xyt}. One observes in both figures that the numerical results of the
GL-5 method again agree very well with the analytical results.

Before giving a more quantitative analysis of the full deviations including the ERDs by using different methods, we comment on
the accuracy in variances and covariances, Eqs. (\ref{eqvarr}) and (\ref{eqcov}), which will help to better understand the
deviations in full trajectories. In deriving Eqs. (\ref{eqvarr}) and (\ref{eqcov}), we used the assumption that the
deterministic force is an explicit function of time only, and we truncated the Taylor series representation for that force.
However, in many realistic problems and in the models such as the BHO, the deterministic force depends explicitly on the
particle position only. While we have seen in the previous sub-section that this assumption can provide a sufficiently accurate
description of the means of the position and velocity for a BHO, a question remains as to how does this assumption affect the
calculation of variances and covariances, i.e. the explicitly random part of displacements.

\begin{figure}
\centering
\includegraphics[trim=80mm 10mm 50mm 10mm, width=0.7\textwidth]{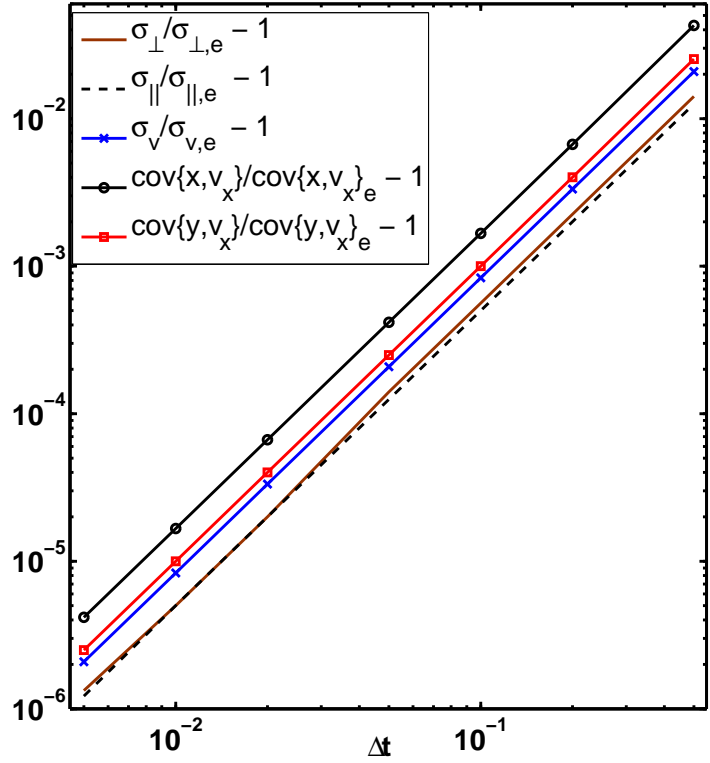}
\caption{(Color online) Deviations of variances and covariances from the corresponding analytical solutions \cite{Jimenez2008}
versus the time step size $\Delta t$, for $\omega_{0}=1/\sqrt{2}$, $\gamma=0.02$ and $\Omega=0.5$.} \label{fig_dev_var_cov}
\end{figure}

This is addressed in Fig.\ \ref{fig_dev_var_cov} which depicts relative deviations of the variances and covariances computed by
the GL-5 method from the corresponding analytical results [Eqs. (26)-(34) in \cite{Jimenez2008}] versus the time step size
$\Delta t$ for $\gamma=0.02$ and $\Omega=0.5$. (Note that, in the specific case of zero magnetic field, the analytical results
of \cite{Jimenez2008} are identical to those of Refs. \cite{Chandrasekhar1943,Lemons2002}.) All deviations are seen to increase
in Fig.\ \ref{fig_dev_var_cov} with increasing $\Delta t$. In the log-log scale, all results form a cluster of parallel straight
lines with a slope of about $2.5$, indicating an apparent order of approximately $2.5$. Judging by both the magnitude of
deviations and by the apparent order, the accuracy for variances and covariances displayed in Fig.\ \ref{fig_dev_var_cov} is
much lower than the accuracy for the means shown in Fig. \ref{fig_dev_vs_dt}. One would expect that the total accuracy of an
algorithm will be largely determined by the accuracy of that part of the algorithm which has the lowest accuracy.

\begin{figure}
\centering
\includegraphics[trim=0 10mm 0 0mm, width=0.9\textwidth]{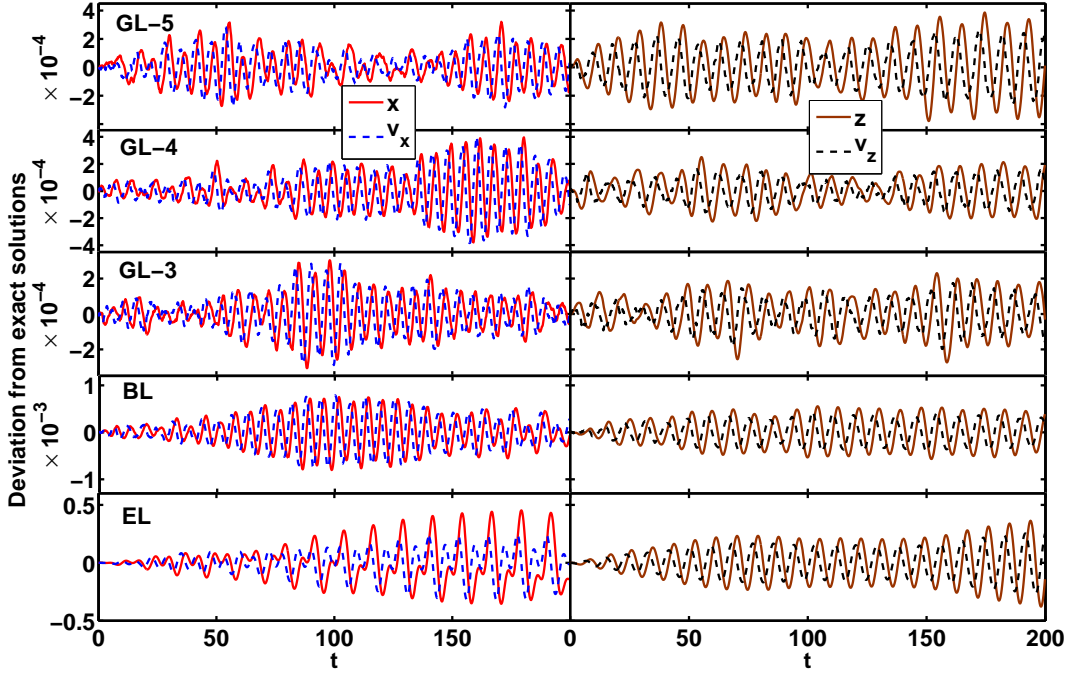}
\caption{(Color online) Deviations of $x$ and $v_x$ (left), and $z$ and $v_z$ (right) from the corresponding analytical
solutions \cite{Jimenez2008} using different methods (from top to bottom: GL-5, GL-4, GL-3, BL and EL) in $200$ time units, with
$\gamma=0.02$, $\omega_{0}=1/\sqrt{2}$, $\Omega=0.5$ and $\Delta t=0.02$. Initial conditions are the same as in Fig.\
\ref{fig_mean_xyz}. Note the magnitude of the errors.} \label{fig_dev_all_xz}
\end{figure}

As before, we next carry out simulations over certain periods and record full deviations (means plus ERDs) of the position and
velocity from the corresponding analytical results \cite{Jimenez2008}. All results are assembled in Fig.\ \ref{fig_dev_all_xz}
for different methods in $200$ time units, with $\Delta t=0.02$, $\gamma=0.02$ and $\Omega=0.5$. One sees that, for all
Gear-like methods the amplitude of the full deviation is about $10^{-4}$, which coincides with the corresponding deviation of
cov\{$x,v_x$\} shown in Fig.\ \ref{fig_dev_var_cov}. This indicates that, for the Gear-like methods, the errors in calculation
come mainly from the evaluation of variances and covariances, that is, from the addition of the ERDs. On the other hand, for the
Beeman-like method and particularly for the Euler-like method, the magnitudes of the full deviations are larger than those of
the deviations of variances and covariances shown in Fig.\ \ref{fig_dev_var_cov}. This implies that the errors introduced by
addition of the ERDs might have been amplified during the calculation of the means. All in all, the dramatic differences in
deviations in the $x$ and $z$ directions seen between different methods in Figs.\ \ref{fig_dev_mean_x}, \ref{fig_dev_mean_z} and
\ref{fig_dev_vs_Omega} have disappeared when ERDs are added.

\begin{figure}
\centering
\includegraphics[trim=30mm 10mm 40mm 0mm, width=0.8\textwidth]{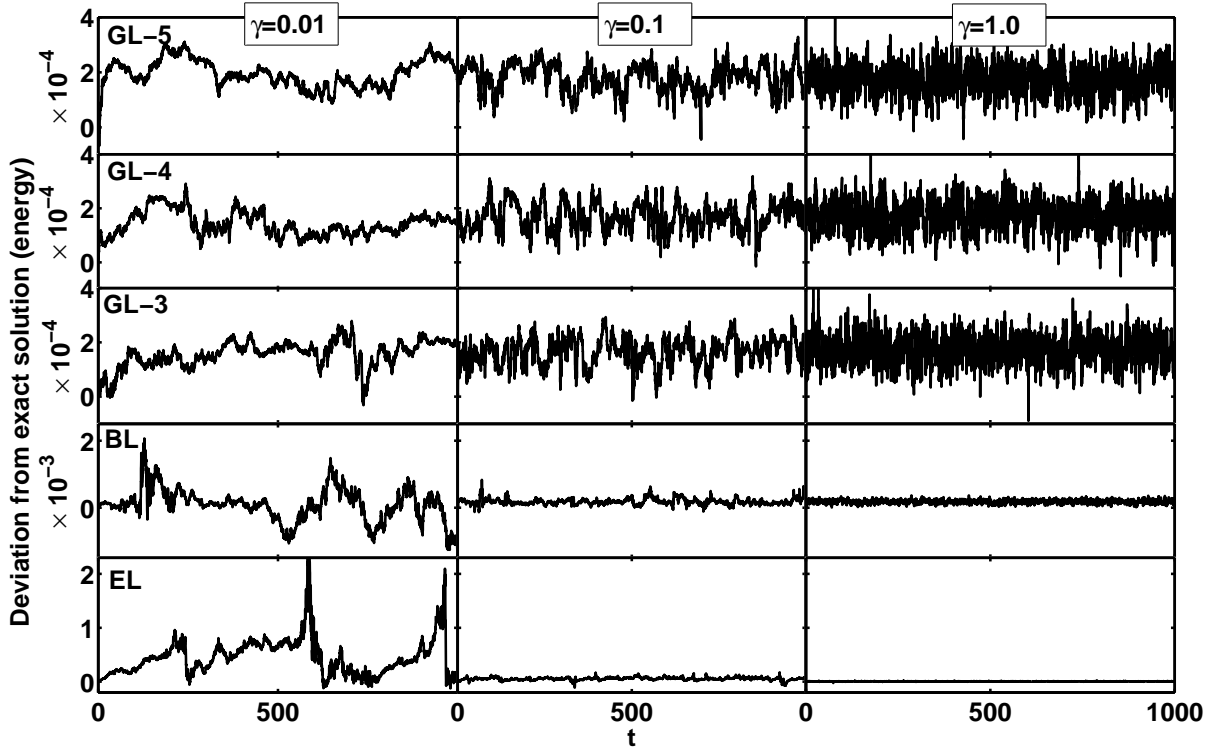}
\caption{ Deviations of the total energy from its exact counterpart \cite{Jimenez2008}, i.e., $E(t)/E_{e}(t)-1$, using different
methods (from top to bottom: GL-5, GL-4, GL-3, BL and EL) for different damping rates $\gamma$ in $1000$ time units, with
$\omega_{0}=1/\sqrt{2}$, $\Omega=0.5$ and $\Delta t=0.02$. Initial conditions are the same as in Fig.\ \ref{fig_mean_xyz}. Note
the magnitude of the errors.} \label{fig_dev_E_gamma}
\end{figure}

From now on, we shall examine mostly the behavior of simulations by monitoring the quantity
\begin{equation}
E(t)=v^{2}(t) + \omega_0^2 r^{2}(t), \label{eqenergy}
\end{equation}
where $v^{2}=v^{2}_{x}+v^{2}_{y}+v^{2}_{z}$ and $r^{2}=x^2+y^2+z^2$, which is defined to be proportional to the total energy of
the Brownian oscillator at time $t$ and, as such, it contains inaccuracies in both the position and velocity. Note that, because
of coupling with the medium through damping and Brownian acceleration, this energy is no longer a conserved quantity. In order
to examine the energy conservation performance of our numerical methods, we normalize $E(t)$ by its ``exact'' counterpart
$E_{e}(t)$, which is simply obtained from Eq.\ (\ref{eqenergy}) by substituting the explicit analytical expressions for the
position and velocity \cite{Jimenez2008}. The resultant ratio $E(t)/E_{e}(t)$ should be then a conserved quantity in a
simulation with the expected value of unity.

Figure \ref{fig_dev_E_gamma} displays the relative deviation of energy from its exact counterpart \cite{Jimenez2008}, i.e.,
$E(t)/E_{e}(t)-1$, calculated by using different methods for several damping rates, with $\Delta t=0.02$ and $\Omega=0.5$. To
check the long time stability of these methods, a longer time scale of $1000$ time units is adopted here. One sees that, for the
Gear-like methods, the amplitude of deviation is quite similar to that of the position and velocity in Fig.\
\ref{fig_dev_all_xz}. Also, the amplitude does not change appreciably with the damping rate, although the frequency of the noise
changes dramatically as the collision frequency, i.e., the damping rate, increases from $\gamma=0.01$ to $1.0$. However, the
situation for the Beeman-like and Euler-like methods is quite different. For the former, the amplitude of the deviation is
approximately one order higher than that of the Gear-like methods at $\gamma=0.01$, but it decreases dramatically when $\gamma$
increases, and reaches almost the same level as that of the Gear-like methods at about $\gamma=1.0$. For the latter, the
amplitude of the deviation is about $1$ at $\gamma=0.01$, and is therefore comparable to the value of energy itself, indicating
that this method is not stable under these conditions. However, it also decreases with increasing damping rate.  So, it is
obvious that finite damping $\gamma$ actually stabilizes Beeman-like and Euler-like methods. This is not surprising at all,
because the damping could also diminish errors inherited from a previous step during simulation. Indeed, previous studies
\cite{Izaguirre2001} have demonstrated a possibility of stabilizing the MD simulation by introducing a small, but finite
damping.

\begin{figure}
\centering
\includegraphics[trim=50mm 40mm 50mm 0mm, width=0.7\textwidth]{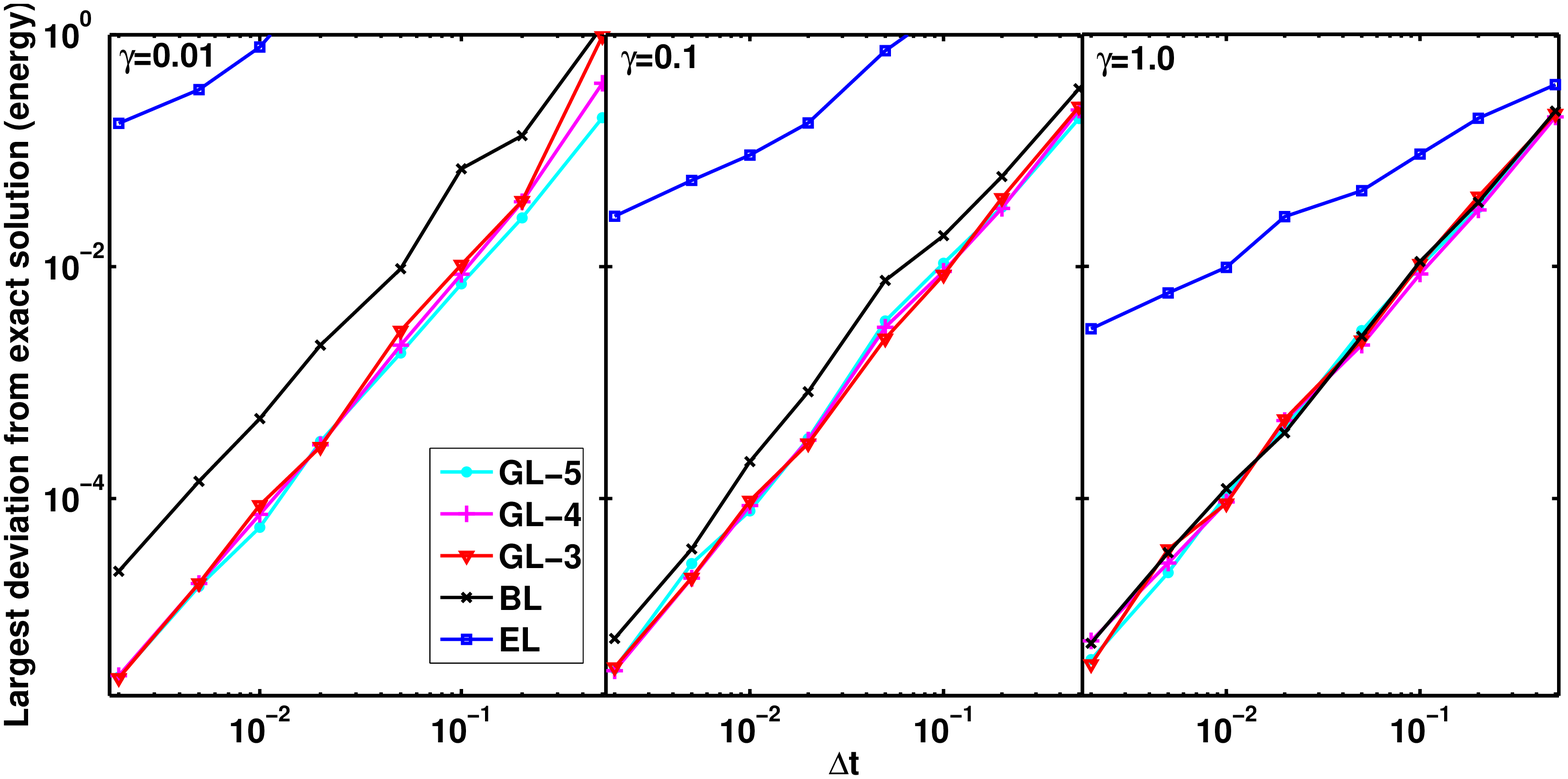}
\caption{(Color online) Largest deviation in total energy from its exact counterpart \cite{Jimenez2008} in $1000$ time units
versus $\Delta t$ with $\omega_{0}=1/\sqrt{2}$ and $\Omega=0.5$ for different methods and different $\gamma$ values. Initial
conditions are the same as those in Fig.\ \ref{fig_mean_xyz}.} \label{fig_dev_E_T1000}
\end{figure}

A more quantitative analysis of these results is shown in Fig.\ \ref{fig_dev_E_T1000}, in which the largest deviations in energy
during $1000$ time units are plotted versus the time step size $\Delta t$ for different $\gamma$ values. One can see that the
deviation of the Euler-like method is always the highest, but it decreases with increasing $\gamma$. Its magnitude suggests that
this method should only be used for simulating systems with damping rate larger than $1.0$ and with very small time steps. As
for the Gear-like methods, judging from the magnitude of deviations and the apparent order of around $2.5$, their accuracy has
reached the limit determined by the accuracy of deviations in variances and covariances shown in Fig.\ \ref{fig_dev_var_cov}.
Therefore, their deviations in the energy remain very close to each other, and they are nearly independent of $\gamma$. The
performance of the Beeman-like method is in between the Euler-like and Gear-like methods for very small $\gamma$, but it reaches
the same level as the Gear-like methods for large damping rates, such as $\gamma=1.0$.

\section{Conclusions}\label{Sec_con}
We have presented several new algorithms for studying Brownian dynamics of charged particles in an external magnetic field. All
these methods were tested by comparison with the available analytical results for a three-dimensional,
Brownian-harmonic-oscillator model in the presence of an external magnetic field \cite{Jimenez2008}. It was found that the
Gear-like method generally has the best performance in terms of accuracy, long time stability, and energy drift in a wide range
of damping rates, and especially in the low-damping limit. Therefore, the Gear-like method should be highly recommended when
studying systems with very low-damping and/or when using the BD method as a thermostat \cite{Ceriotti2009} in a MD simulation.
The Beeman-like method can also cover a wide range of damping rates with reasonably good accuracy and with negligible energy
drift. It should be recommended for simulating systems with intermediate damping rates. The Euler-like method, as can be
expected, has the poorest performance and can be used with confidence only in simulating over-damped systems, such as colloidal
suspensions and/or polymeric fluids. Further detailed tests based on applications to magnetized complex plasmas will be
presented elsewhere \cite{Comingup}.

We note that, besides applications in plasma physics, our numerical method could be also of interest in numerical studies of
some stochastic processes in statistical physics \cite{Jimenez2008,Jimenez2008b,Jimenez,Jayannavar2007,Roy2008}, since we have
actually tested here the recently developed analytical model for a Brownian-harmonic-oscillator in the presence of a magnetic
field.

\begin{acknowledgments}
L.J.H. acknowledges support from Alexander von Humboldt Foundation. Work at CAU is supported by DFG within SFB-TR24/A2. Z.L.M.
acknowledges support from NSERC.
\end{acknowledgments}

\end{document}